\documentclass[aps,prl,letterpaper,twocolumn,showpacs]{revtex4}%
\usepackage{amsfonts}
\usepackage[T2A]{fontenc}
\usepackage[cp1251]{inputenc}
\usepackage{amsmath}
\usepackage{amssymb}
\usepackage[russian,english]{babel}
\usepackage{graphicx}
\begin{document}
\author{Yu.Kagan, D.L.Kovrizhin, and L.A.Maksimov}
\affiliation{"RRC Kurchatov Institute", Kurchatov Square 1, 123182 Moscow, Russia}
\title{The anomalous tunneling of Bose-condensate excitations}

\begin{abstract}
We discuss the tunneling of phonon excitations across a potential barrier
separating two condensate bulks. It is shown that the strong barrier proves to
be transparent for the excitations at low energy $\varepsilon.$ Moreover, the
transmission is reduced with increasing $\varepsilon$ in contrast to the
standard dependence. This anomalous behavior is due to an existence of the
quasiresonance interaction. The origin of this interaction is a result of the
formation of the special well determined by the density distribution of
condensate in the vicinity of a high barrier.

\end{abstract}
\date{15 October 2002}
\maketitle

The investigation of the tunneling phenomena in the Bose-condensed systems
becomes one of the most evolving trends in the BEC studies of ultracold atomic
gases, (see e.g., [1] and references therein). One of the interesting aspects
in this field is associated with the problem of the tunneling of collective
excitations possessing specific features in the condensed phase. The
condensate moving at velocity $v$ smaller than some critical velocity $v_{c}$
transmits across a potential barrier without scattering and reflection.
Excitations of a condensate at rest, in particular, sound excitations should
experience the tunneling transmission together with the reflection. It is
significant that the motion of excitations takes place at the background of
the inhomogeneous built-up distribution of condensate density $g\left(
x\right)  $. The inevitable reduction of $g\left(  x\right)  $ near the
barrier results in the appearance of a potential well. For the sufficiently
high barrier $V\gg\mu$ where $\mu$ is the chemical potential, the tunneling of
excitations under these conditions, as we will see below, demonstrates an
anomalous character. On the one hand, the barrier proves to be transparent for
phonon excitations within a limited range of low energies. On the other hand,
the coefficient of tunneling transmission $T$ reduces as the energy of
excitations $\varepsilon$ grows. This is in contrast to the typical growth of
$T$ with $\varepsilon.$ Such anomalous behavior to a great extent is
associated with the appearance of a quasi-bound state in the continuous
spectrum. The formation of the quasi-bound state is connected with the
specific behavior of function $g\left(  x\right)  $ in the vicinity of the
barrier. Provided the size $L$ of the condensate is much larger than the
correlation length $\xi$ and the barrier is high, the dependence
\thinspace$g\left(  x\right)  $ in essential degree has an universal
character. As a result, anomalous picture of tunneling acquires a relatively
universal character as well. In order to make the exposition ultimate clear,
we consider one-dimensional symmetrical configuration, namely, two identical
condensates separated by a rectangular barrier with height $V_{0}\gg\mu$. The
general equation for the Heisenberg field operator of particles $\hat{\Psi
}\left(  x,t\right)  $ has the form%
\begin{equation}
i\hbar\frac{\partial\hat{\Psi}}{\partial t}=\left(  -\frac{\hbar^{2}}%
{2m}\frac{\partial^{2}}{\partial x^{2}}-\mu\right)  \hat{\Psi}+V\left(
x\right)  \hat{\Psi}+U\hat{\Psi}^{+}\hat{\Psi}\hat{\Psi}.\label{1}%
\end{equation}
Hereafter it is implicitly assumed that the transverse size of the system is
sufficiently large. Accordingly a collision of atoms has a three-dimensional
character and, as usual, $U=4\pi\hbar^{2}a/m$ where $a$ is the scattering
length. We consider only the case $a>0.$

We start from the consideration of the ground state, assuming temperature
$T=0$. In this case the operator $\hat{\Psi}$ can be replaced with the
macroscopic condensate wave function $\Psi\left(  x\right)  .$ The equation
for $\Psi\left(  x\right)  $ in the dimensionless form can be written as%
\begin{equation}
i\frac{\partial\bar{\Psi}}{\partial\bar{t}}=\left(  -\frac{1}{2}%
\frac{\partial^{2}\bar{\Psi}}{\partial\bar{x}^{2}}-1\right)  \bar{\Psi}%
+\bar{V}\left(  \bar{x}\right)  \bar{\Psi}+\left\vert \bar{\Psi}\right\vert
^{2}\bar{\Psi}. \label{2}%
\end{equation}
Here we introduced the following notations%
\begin{equation}%
\begin{array}
[c]{llll}%
\bar{x}=x/\xi, & \bar{t}=\mu t/\hbar, & \bar{\Psi}=\Psi\left(  U/\mu\right)
^{\frac{1}{2}}, & \bar{V}=V/\mu.
\end{array}
\label{3}%
\end{equation}
Further we omit bar for all variables in Eq.(\ref{2}).

Let us find a stationary solution $\Psi_{0}\left(  x\right)  $ of
Eq.(\ref{2}), assuming that $\Psi_{0}\left(  x\right)  =1$ for $\left\vert
x\right\vert \gg1.$ Outside barrier Eq.(\ref{2}) has the known static solution%
\begin{equation}
\Psi_{0}\left(  x\right)  =\tanh\left(  \left\vert x\right\vert +x_{0}\right)
. \label{4}%
\end{equation}
Under the barrier we can find the first integral of Eq.(\ref{2}), taking into
account the condition $\Psi_{0}^{\prime}\left(  0\right)  =0$%
\begin{equation}
\Psi_{0}^{\prime}\left(  x\right)  =\sqrt{\left(  \Psi_{0}^{2}-b^{2}\right)
\left(  \Psi_{0}^{2}+\varkappa_{0}^{2}\right)  }. \label{e5}%
\end{equation}
Here $b^{2}=\Psi^{2}\left(  0\right)  ,\ \varkappa_{0}^{2}=2\left(
V_{0}-1\right)  +b^{2}.$ The general solution takes the form%
\begin{equation}
\Psi_{0}\left(  x\right)  =\frac{b}{\operatorname{cn}\left(  \sqrt
{b^{2}+\varkappa_{0}^{2}}x\right)  }, \label{e6}%
\end{equation}
where $\operatorname{cn}u$ is the Jacobi elliptic function (see, e.g.
\cite{ryzhik}).

In the case of strong barrier when $\varkappa_{0}\gg1$ and $\varkappa_{0}d>1$
($d=2a$ is the barrier width), one can neglect the nonlinearity related to the
interparticle interaction. Equating solutions (\ref{4}),(\ref{e6}) and their
derivatives at the boundary of the barrier $\left\vert x\right\vert =a,$ we
derive%
\begin{equation}
\Psi_{0}\left(  x\right)  \cong\left(  \varkappa_{0}\sinh\varkappa
_{0}a\right)  ^{-1}\cosh\varkappa x,\ \Psi_{0}\left(  a\right)  \approx
\varkappa_{0}^{-1}\ll1. \label{e7}%
\end{equation}

Let us now consider the tunneling of excitations propagating at the background
of the distribution of the condensate density. These excitations can be found
as oscillations of the classical field of a condensate (see,
e.g.\cite{dalfovo}). For this purpose, let us represent function $\Psi$ in
Eq.(\ref{2}) as $\Psi=\Psi_{0}+\Psi^{\prime}$ and linearize the equation in
$\Psi^{\prime}$%
\begin{equation}
i\frac{\partial\Psi^{\prime}}{\partial t}=\hat{h}\Psi^{\prime}+g\left(
x\right)  \Psi^{\prime\ast}, \label{e10}%
\end{equation}%
\begin{equation}%
\begin{array}
[c]{ll}%
\hat{h}=-\dfrac{1}{2}\dfrac{d^{2}}{dx^{2}}+V\left(  x\right)  +2g\left(
x\right)  -1, & g\left(  x\right)  =\Psi_{0}^{2}\left(  x\right)  .
\end{array}
\label{e11}%
\end{equation}

For excitations with energy $\varepsilon,$ we will seek the solution of a set
of equations (\ref{e10}) for $\Psi^{\prime}$ and $\Psi^{\prime\ast}$ in the
form \cite{dalfovo}%
\begin{equation}
\Psi^{\prime}=u\left(  x\right)  e^{-i\varepsilon t}-v^{\ast}\left(  x\right)
e^{i\varepsilon t}.\label{e12}%
\end{equation}
Then we have%
\[
\left(  \hat{h}-\varepsilon\right)  u-gv=0,\ -gu+\left(  \hat{h}%
+\varepsilon\right)  v=0.
\]
Introducing notations%
\begin{equation}%
\begin{array}
[c]{ll}%
S=u+v,\ \ \  & G=u-v
\end{array}
\label{e13}%
\end{equation}
and combining these equations, we arrive at fourth-order equation%
\begin{equation}
\left(  \hat{h}+g\right)  \left(  \hat{h}-g\right)  S=\varepsilon
^{2}S.\label{e14}%
\end{equation}
If the solution of $S\left(  x\right)  $ is known, one can readily find
function $G\left(  x\right)  $%
\begin{equation}
\varepsilon G=\left(  \hat{h}-g\right)  S.\label{e15}%
\end{equation}
Far from the barrier at $\left\vert x\right\vert \gg1,$ we have $g\left(
x\right)  \rightarrow1.$ The solutions of Eq.(\ref{e14}) in this uniform
region are sought as $S\sim\exp\left(  ikx\right)  .$ As a result, we find
four roots%
\begin{equation}%
\begin{array}
[c]{ll}%
k_{1,2}=\pm k, & k=\sqrt{2}\left(  \sqrt{1+\varepsilon^{2}}-1\right)
^{1/2},\\
k_{3,4}=\mp iq, & q=\sqrt{2}\left(  \sqrt{1+\varepsilon^{2}}+1\right)  ^{1/2}.
\end{array}
\label{e16}%
\end{equation}
The first two roots corresponds to the Bogoliubov spectrum (in the dimensional
variables $k\rightarrow k\xi$ and $\varepsilon\rightarrow\varepsilon/\mu$
respectively). The solutions corresponding to roots $k_{3,4}$ do not
contribute into the asymptote of the general solution at $\left\vert
x\right\vert \gg1$ but prove to be significant in the regions close to the
barrier. For $\varepsilon\ll1,$ we have%
\begin{equation}
k=\varepsilon,\ \ \text{ }q=2,\label{e17}%
\end{equation}
neglecting the terms of the order of $\varepsilon^{2}.$

Outside barrier the general solution can be represented as a linear
superposition of basic solutions $S_{i}\left(  x\right)  $ of equation
(\ref{e14}) corresponding to the roots given in (\ref{e16}). Let excitations
fall from the left hand site. Then, omitting the divergent components, we have%
\begin{equation}%
\begin{array}
[c]{ll}%
S\left(  x\right)  =AS_{1}\left(  x\right)  +BS_{2}\left(  x\right)
+CS_{3}\left(  x\right)  , & x<-a\\
S\left(  x\right)  =DS_{4}\left(  x\right)  +FS_{5}\left(  x\right)  , & x>a.
\end{array}
\label{e18}%
\end{equation}
Here the functions $S_{n}\left(  x\right)  $ for fixed $\varepsilon$ have an
universal form%
\begin{equation}%
\begin{array}
[c]{l}%
S_{n}\left(  x\right)  =e^{ik_{n}x}\left(  \tanh\left(  -x+x_{0}\right)
+\dfrac{ik_{n}}{2}\right)  ,\ n=1,2,3\\
S_{4,5}\left(  x\right)  =S_{2,3}\left(  -x\right)  .
\end{array}
\label{e19}%
\end{equation}
Function $G\left(  x\right)  $ can be found from (\ref{e15}) with conserving
the same set of the coefficients to be determined as in (\ref{e18}).

Consider now the underbarrier region. Since $g\left(  x\right)  \leqslant$
$\varkappa_{0}^{-2}\ll1,$ we first find the solution of Eq.(\ref{e14})
omitting $g\left(  x\right)  $. In this case Eq.(\ref{e14}) can be rewritten
as%
\begin{equation}
\left(  -\frac{1}{2}\frac{d^{2}}{dx^{2}}+V-1\right)  ^{2}S=\varepsilon
^{2}S.\label{e20}%
\end{equation}
Inserting the solution as $S\sim\exp\left(  -\varkappa x\right)  $ into the
equation, we again find four roots%
\begin{equation}%
\begin{array}
[c]{ccc}%
\varkappa_{1,2}=\pm\varkappa_{+}, & \varkappa_{3,4}=\pm\varkappa_{-}, &
\varkappa_{\pm}=\varkappa_{0}\sqrt{1\pm2\varepsilon/\varkappa_{0}^{2}}.
\end{array}
\label{e21}%
\end{equation}
Accordingly, the general solution for the underbarrier region reads%
\[
S_{B}=Ke^{\varkappa_{+}x}+Le^{-\varkappa_{+}x}+Me^{\varkappa_{-}%
x}+Ne^{-\varkappa_{-}x}.
\]
Substituting this expression into (\ref{e15}) and involving (\ref{e21}), we
arrive at%
\[
G_{B}=-\left(  Ke^{\varkappa_{+}x}+Le^{-\varkappa_{+}x}\right)  +Me^{\varkappa
_{-}x}+Ne^{-\varkappa_{-}x}.
\]
Now we can employ that $\varepsilon/\varkappa_{0}^{2}\ll1$ and, supposing
$\varepsilon a/\varkappa_{0}\ll1,$ one can replace $\varkappa_{\pm}$ with
$\varkappa_{0}.$ As a result, we have%
\begin{equation}%
\begin{array}
[c]{l}%
S_{B}=\left(  M+K\right)  e^{\varkappa_{0}x}+\left(  N+L\right)
e^{-\varkappa_{0}x},\\
G_{B}=\left(  M-K\right)  e^{\varkappa_{0}x}+\left(  N-L\right)
e^{-\varkappa_{0}x}.
\end{array}
\label{e22}%
\end{equation}
\bigskip\ The boundary conditions at $x=\pm a$ yield eighth equations for the
determination of all coefficients in (\ref{e18}) and (\ref{e22})%
\begin{equation}%
\begin{array}
[c]{cc}%
S\left(  \pm a\right)  =S_{B}\left(  \pm a\right)  \ \left(  \alpha\right)
, & G\left(  \pm a\right)  =G_{B}\left(  \pm a\right)  \ \left(  \beta\right)
,\\
\left.  \dfrac{dS}{dx}\right\vert _{\pm a}=\left.  \dfrac{dS_{B}}%
{dx}\right\vert _{\pm a}\ \left(  \gamma\right)  , & \left.  \dfrac{dG}%
{dx}\right\vert _{\pm a}=\left.  \dfrac{dG_{B}}{dx}\right\vert _{\pm
a}\ \left(  \delta\right)  .
\end{array}
\label{e23}%
\end{equation}

In order to find the solution in the explicit form, we restrict ourselves by
considering the tunneling and reflection of collective excitations in the
sound region corresponding to the roots (\ref{e17}). In addition, we simplify
a set of equations omitting quadratic terms like $k^{2},$ $k\zeta^{1},$
$\zeta^{2}$ ($\zeta=\varkappa_{0}^{-1}\ll1$) and involving the smallness of
coefficients $C$ and$\ F$ of the order of ($k,\zeta$) with respect to the
other coefficients in Eq.(\ref{e18}). This simplification yields the
equalities $M\approx K$ and $N\approx L$ resulting from Eq.(\ref{e23}$\delta
$). Together with the boundary conditions for $G$ (\ref{e23}$\beta$) this
gives%
\begin{equation}%
\begin{array}
[c]{ll}%
\bar{F}=-\bar{D}\dfrac{ik}{2}, & \bar{C}=\bar{A}\dfrac{ik}{2}-\bar{B}%
\dfrac{ik}{2}.
\end{array}
\label{e24}%
\end{equation}
Here
\[%
\begin{array}
[c]{lll}%
\bar{F}=Fe^{-2a}, & \bar{C}=Ce^{-2a}, & \bar{B}=Be^{ika},\\
\bar{D}=De^{ika}, & \bar{A}=Ae^{-ika}. &
\end{array}
\]
The remaining set of four equations can be solved straightforwardly%
\begin{equation}%
\begin{array}
[c]{cc}%
\bar{B}=\frac{\bar{A}}{Z}(-k^{2}+\left(  k^{2}+4\zeta^{2}\right)
e^{-4\varkappa_{0}a}), & \bar{D}=\frac{\bar{A}}{Z}4ik\zeta e^{-2\varkappa
_{0}a},\\
K=\frac{\bar{A}}{Z}\zeta k\left(  k+2i\zeta\right)  e^{-3\varkappa_{0}a}, &
L=\frac{\bar{A}}{Z}\zeta k^{2}e^{-\varkappa_{0}a},\\
Z=k^{2}-\left(  k+2i\zeta\right)  ^{2}e^{-4\varkappa_{0}a}. &
\end{array}
\label{e25}%
\end{equation}
As a result, returning to dimensional variables, we find for the tunneling
transmission of the excitations%
\begin{equation}
T=\left\vert \frac{D}{A}\right\vert ^{2}\approx\frac{16\left(  k/\varkappa
_{0}\right)  ^{2}e^{-4\varkappa_{0}a}}{\left[  \left(  k\xi\right)
^{2}+\left(  4\left(  \varkappa_{0}\xi\right)  ^{-2}-\left(  k\xi\right)
^{2}\right)  e^{-4\varkappa_{0}a}\right]  ^{2}}. \label{e26}%
\end{equation}
Even for a wide barrier the transmission coefficient demonstrates practically
the full transparency at%
\[
k=k_{\ast}\approx\frac{1}{\xi}\frac{2}{\varkappa_{0}\xi}e^{-2\varkappa_{0}a}.
\]
For $k\gg k_{\ast}$ and still $k\xi\ll1,$ the transmission coefficient falls
with increasing $k$%
\begin{equation}
T\approx\frac{16e^{-4\varkappa_{0}a}}{\left(  \varkappa_{0}\xi\right)
^{2}\left(  k\xi\right)  ^{2}}. \label{e27}%
\end{equation}
This result is in a drastic contrast with the typical enhancement of $T$ with
the growth of $\ k.$

Let us consider a more accurate solution for the underbarrier region,
involving the condensate density $g\left(  x\right)  $ in Eqs.(\ref{e14}%
),(\ref{e15}). The explicit form for $g\left(  x\right)  $ can be obtained
from the results (\ref{e6})-(\ref{e7}) for $\Psi_{0}\left(  x\right)  .$ The
numerical calculation should be performed in order to find the solution of
Eqs.(\ref{e14}),(\ref{e15}) for the underbarrier region. The general solution
implies again the use of the boundary conditions (\ref{e23}). The results of
calculations for the dependence $T$ on $k$ are presented in Fig.1. In these
figures the function $T\left(  k\right)  ,$ obtained with ignoring $\ g\left(
x\right)  $ in the underbarrier region, is shown by the dashed lines. The
exact solution shows again the existence of the full transparency. The only
difference is a shift of the maximum of $T$ to $k=0$. With increasing $k$, in
particular, in the region where the asymptotic expression (\ref{e27}) is
correct, the both curves coincide. We intentionally choose quite different
sets of the parameters in Fig.1(a,b) in order to demonstrate rather universal
character of the $T\left(  k\right)  $ behavior. In fact, we see only the
change of the scale of $k$-region where the transparency is high.%
\begin{figure}
\includegraphics{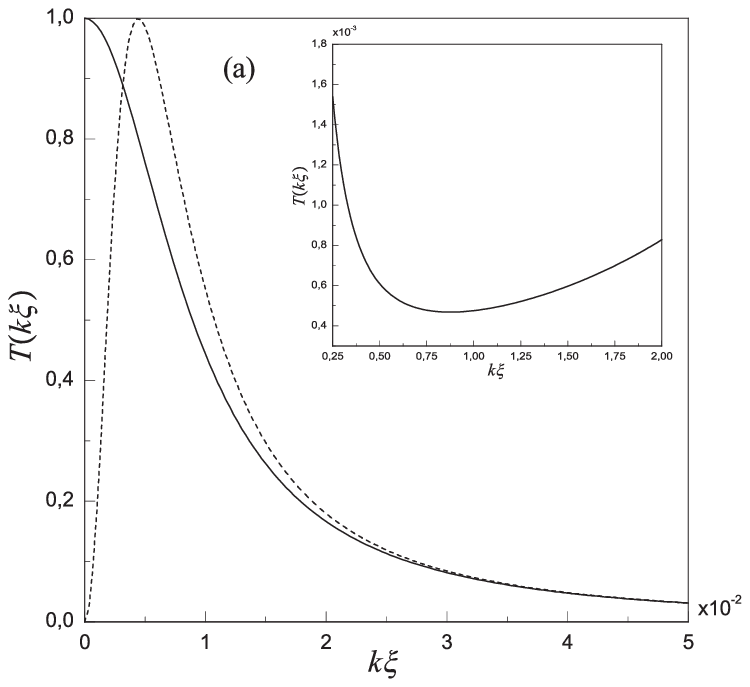}
\includegraphics{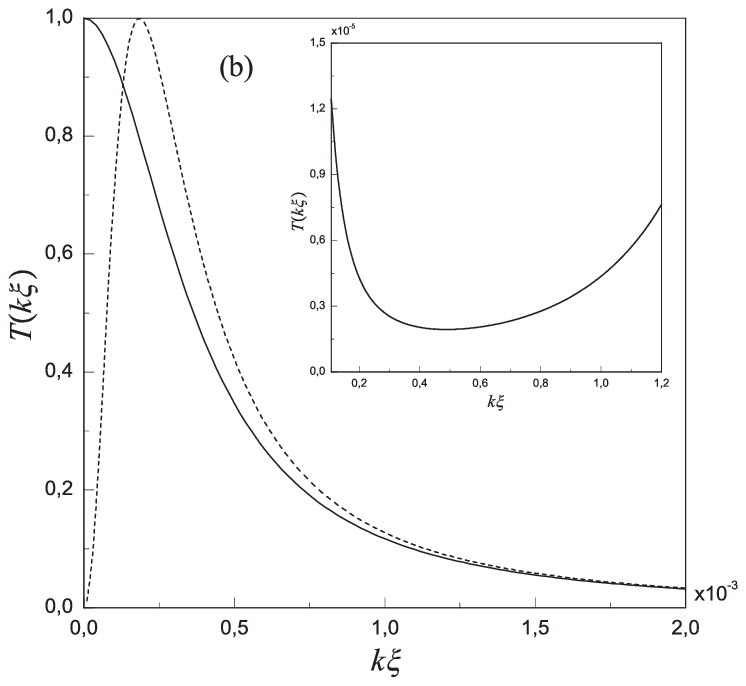}
\caption{\label{fig1} The coefficient of tunneling transmission $T$,
as a function of $k\xi$ for two sets of parameters:
(a) $\varkappa_0\xi=8, \varkappa_0 d=4$;
(b) $\varkappa_0\xi=3, \varkappa_0 d=8$.
Dashed line corresponds to $T(k\xi)$ obtained with neglecting  $g(x)$
in the underbarrier region.}
\end{figure}%

The anomalous behavior of the tunneling transmission of excitations has an
interesting origin. In the vicinity of the barrier the condensate density
decreases as $\tanh^{2}\left[  \left(  \left\vert x\right\vert +x_{0}\right)
/\xi\right]  $, producing potential wells for the excitations on the both
sides of a barrier. In the case of a strong barrier the ratio $\left(
\left\vert x\right\vert +x_{0}\right)  /\xi\ll1$. In these conditions a
virtual resonance level in the continuous spectrum close to $\varepsilon=0$
appears. This results in the drastic change of the tunneling picture. As $k$
increases, the quasi-resonance scattering decreases entailing the unusual
reduction of $T\left(  k\right)  $. At $k\xi\sim1$ the role of virtual level
reduces and the normal growth of $T$ with $\ k$ should be restored. We can see
it from the inserts in Fig.1(a,b) where the dependence $T\left(  k\right)  $
is plotted just for this interval of $k.$ Note that in all cases the anomalous
tunneling can be revealed considering the reflection coefficient%
\begin{equation}
R=\left\vert \frac{B}{A}\right\vert ^{2}=1-T. \label{e28}%
\end{equation}

Treating the tunneling of phonon excitations, it is interesting to trace the
formation of energy flux $Q$. In the general case this flux can be found
employing the local form of the energy conservation law. Provided the
Hamiltonian is represented as%
\[
H=\int E\left(  r\right)  d^{3}r,
\]%
\begin{equation}
E\left(  r\right)  =\frac{\hbar^{2}}{2m}\nabla\hat{\Psi}^{+}\nabla\hat{\Psi
}+\hat{\Psi}^{+}\left(  V-\mu\right)  \hat{\Psi}+\frac{U}{2}\hat{\Psi}^{+}%
\hat{\Psi}^{+}\hat{\Psi}\hat{\Psi}. \label{e30}%
\end{equation}
then%
\begin{equation}
\frac{\partial E}{\partial t}=-\operatorname{div}Q \label{30}%
\end{equation}
Using Eqs.(\ref{e30}) and (\ref{1}), we find%
\begin{equation}
Q=-\frac{\hbar^{2}}{m}\operatorname{Re}\left(  \frac{\partial\hat{\Psi}^{+}%
}{\partial t}\frac{\partial\hat{\Psi}}{\partial x}\right)  . \label{e31}%
\end{equation}
Assuming as before that the condensate is at rest, we consider the energy
transfer by excitations. Treating them as oscillations of the classic field of
a condensate, we have from (\ref{e31}) for the flux averaged over the time%
\begin{equation}
\left\langle Q\right\rangle =n_{0}\frac{\mu^{3/2}}{m^{1/2}}\left\langle
\bar{Q}\right\rangle ,\ \ \left\langle \bar{Q}\right\rangle
=-\operatorname{Re}\left\langle \frac{\partial\Psi^{\prime\ast}}{\partial
t}\frac{\partial\Psi^{\prime}}{\partial x}\right\rangle . \label{e32}%
\end{equation}
Here $n_{0}$ is the total particle density, $\bar{Q}$ is the expression for
the energy flux defined in terms of dimensionless units (\ref{3}). Let us
consider again the incident flux of excitations with energy $\varepsilon.$
Inserting the function $\Psi^{\prime}$ from (\ref{e12}) and using notations
(\ref{e13}), we obtain%

\begin{equation}
\left\langle \bar{Q}\right\rangle =\frac{\varepsilon}{2}\operatorname{Im}%
\left(  S^{\ast}\frac{dS}{dx}+G^{\ast}\frac{dG}{dx}\right)  . \label{e34}%
\end{equation}
Outside the barrier we use the general solution for $S$ in the form
(\ref{e18}) and (\ref{e19}). Taking into account that function $S_{3}$ is real
and $S_{2}\left(  k,x\right)  =S_{1}^{\ast}\left(  k,x\right)  =S_{1}\left(
-k,x\right)  ,$ we find for the region $x<-a$%
\[
\operatorname{Im}\left(  S^{\ast}\frac{dS}{dx}\right)  =\left\vert
A\right\vert ^{2}\left(  1-R\right)  \frac{k}{2}\left(  1+\tanh^{2}\left(
x-x_{0}\right)  \right)  ,
\]
where $R$ is the reflection coefficient (\ref{e28}). According to (\ref{e15})
the contribution into (\ref{e34}) from the function $G$ equals%
\[
\operatorname{Im}\left(  G^{\ast}\frac{dG}{dx}\right)  =\left\vert
A\right\vert ^{2}\left(  1-R\right)  \frac{k}{2}\left(  1-\tanh^{2}\left(
x-x_{0}\right)  \right)  .
\]
A sum of both expressions is independent of $x,$ demonstrating the
conservation of the energy flux of excitations.

Let us find the relation between $\left\vert A\right\vert ^{2}$ and the
physical parameters. Comparing the density of noncondensate particles
$n^{\prime}=\frac{1}{2}n_{0}\left\vert A\right\vert ^{2}$ in the incident
phonon flux with the known Bogoliubov result (see e.g. \cite{lifshitz}) for
the relation between $n^{\prime}$ and phonon density $n(k)$ concentrated in
the $k$-mode $n^{\prime}=(\mu/\hbar c_{0}k)$, we find%
\begin{equation}
\left\vert A\right\vert ^{2}\approx2(\mu/\hbar c_{0}k)(n\left(  k\right)
/n_{0}).\label{e38}%
\end{equation}
Here $c_{0}=\sqrt{\mu/m}$ is the speed of sound defined for the density at
$\left\vert x\right\vert \gg a.$

Returning to the dimensional units, we find for the energy flux (\ref{e31})%
\begin{equation}
\left\langle Q\right\rangle =n\left(  k\right)  c_{0}\varepsilon\left(
1-R\right)  .\label{e39}%
\end{equation}

In order to obtain the flux under the potential barrier in the explicit form,
we neglect again the distribution of the condensate density in this region.
Taking into account the solution (\ref{e22}) and the conditions $M\approx K$
and $N\approx L$, we find from (\ref{e34}) $\left\langle \bar{Q}\right\rangle
=4\varepsilon\varkappa\operatorname{Im}\left(  KL^{\ast}\right)  $. The
coefficients $K$ and $L$ are determined in (\ref{e25}) with $\left\vert
A\right\vert ^{2}$ according to (\ref{e38}). As a result, the energy flux
(\ref{e32}) acquires the value%
\begin{equation}
\left\langle Q\right\rangle =n\left(  k\right)  c_{0}\varepsilon T,\label{e40}%
\end{equation}
where $T$ is the transmission coefficient (\ref{e26}). At $x\geq a$ the flux
has the same value (\ref{e40}). Expressions (\ref{e39}) and (\ref{e40})
demonstrate the constancy of the energy flux in the system (see Eq.(\ref{e28}%
)). Note that this is not the case for a particle flux. One can show that this
flux inevitably connected with the transfer of phonon excitations, is not
conserved and depends on $x.$

\begin{acknowledgments}
This work is supported by INTAS (project 2001-2344), by Netherlands
Organization for Science Research (NWO), and by the Russian Foundation for
Basic Research.
\end{acknowledgments}

\end{document}